\documentstyle[epsfig,letters]{mn1}

\newcommand{\lens}{{\rm lens}}
\newcommand{\gal}{{\rm gal}}

\newcommand{\da}{d_A}
\newcommand{\rad}{r}

\newcommand{\bn}{{\bf \hat{n}}}

\newlength{\tskip}
\newlength{\colwidth}

\newcommand{\beq}{\begin{equation}}
\newcommand{\eeq}{\end{equation}}
\newcommand{\beqa}{\begin{eqnarray}}
\newcommand{\eeqa}{\end{eqnarray}}

\title[Galaxy-Mass Cross Power Spectrum]{Weak Lensing Probes of the Galaxy-Mass Cross Power Spectrum}
\author[Cooray]{Asantha Cooray\\
Division of Physics, Mathematics and Astronomy, California
Institute of Technology, Pasadena, CA 91125.
E-mail: asante@caltech.edu}
\date{\today}
\begin{document}
\maketitle


\begin{abstract}
The angular correlation function of the background shear-foreground galaxy distribution probes the three dimensional cross power spectrum between mass and galaxies. The same cross power spectrum is also probed when foreground galaxy distribution  is cross-correlated with a distribution of background sources disjoint in redshift space. The kernels  that project three dimensional clustering to the two dimensional angular space is different for these two probes. When combined, they  allow a study of the galaxy-mass cross power spectrum 
from linear to non-linear scales.  By inverting the background shear-foreground galaxy correlation function measured by the Sloan Digital Sky Survey, we present a first estimate of the cross power spectrum between mass and galaxies at low redshifts.
\end{abstract}
 

\section{Introduction}

The current and upcoming wide field imaging data, such as from the Sloan Digital Sky Survey, allow detailed studies on the angular clustering of galaxies and quasars (see, e.g., Dodelson et al. 2001; Scranton 2002). In addition to such studies, these data also allow investigations related to the cross clustering between galaxies and mass.  With the so-called galaxy-galaxy lensing (Blandford et al. 1991; Bartelmann \& Schneider 2002 for a recent review), one measures the correlation between foreground galaxies and the background shear surrounding these galaxies. This results in a measuremnt of  the correlation between foreground galaxies and the dark matter distribution traced by these galaxies. Under the assumption that a single galaxy reside in each  dark matter halo, one can use the observed correlation function to constrain some physical properties of halos  (Fischer et al. 2000). Alternatively, more detailed, and increasingly complicated, models can also be introduced such that one takes in to account the fact that more than one galaxy may be present in 
dark matter halos at the high end of the mass function (Guzik \& Seljak 2001, 2002).

In addition to the background shear-foreground galaxy
 correlation function, the cross-correlation
function between  a sample of foreground galaxies and
 background sources, such as quasars, also probe the same galaxy-dark matter cross power spectrum (Moessner et al. 1997).
 This correlation results from the fact that number counts of background galaxies are affected by lensing magnification via the intervening dark matter distribution traced by foreground galaxies. 

In this {\it letter}, we will briefly consider complimentary properties of these two lensing probes of the same galaxy-mass cross power spectrum. The foreground galaxy-background source correlation and the foreground galaxy-background shear correlation  have unique properties in that they probe two different regimes of the galaxy-dark matter cross-power spectrum. We will describe this difference in terms of the kernel involved with the projection of three dimensional clustering to the two dimensional angular space.
Finally,  we will estimate the cross power spectrum between galaxies and mass by inverting the published galaxy-shear correlation function from the Sloan Digital Sky Survey. 
When necessary, we will illustrate our results using the currently favored $\Lambda$CDM cosmology with $\Omega_m=0.35$ and $\Omega_\Lambda=0.65$ and use inputs, such as the redshift distributions necessary for inversions, from published results in the literature.

\section{Shear-Galaxy correlation}
 
As briefly discussed earlier, the shear-galaxy correlation function can be
constructed by correlating tangential shear of background galaxies
surrounding foreground galaxies. The assumption is that these foreground
galaxies trace the mass distribution  responsible for weak lensing
of background sources.  To derive the associated correlation function, it is useful to first consider the relation between  mean tangential shear, 
$\left< \gamma_t(\theta) \right>$,  and convergence:
\begin{equation}
\left< \gamma_t(\theta) \right> = -\frac{\theta}{2} \frac{d
\bar{\kappa}(\theta)}{d \theta}  \, ,
\label{eqn:tangential}
\end{equation}
where $\bar{\kappa}(\theta)$ is the mean convergence within a circular
radius of $\theta$ (Kaiser \& Squires 1993).
The foreground sources are assumed to trace the dark matter distribution and 
one can write  fluctuations in the
foreground source population, $\delta N_f(\bn)$, as
\begin{equation}
\delta N_f(\bn) = \int d\rad W^\gal_f(r) \delta_g(\bn,\rad) \, ,
\label{eqn:foreground}
\end{equation}
where $W^\gal_f(\rad)$ is the normalized distribution of foreground galaxies in radial coordinates with comoving radial distance given by $\rad$, while $\delta_g$ is the fluctuation in the galaxy density field. 

Since the shear, averaged over a circular aperture, is correlated with foreground galaxy positions, one essentially probe the
galaxy-mass correlation. 
We write under the Limber approximation (Limber 1954) using Fourier expansion of equation~(\ref{eqn:foreground}) as
\begin{equation}
\bar{\kappa}(\theta) = \int d\rad W^\lens(\rad)W^\gal_f(\rad)
\int \frac{k dk}{2\pi} P_{g\delta}(k) \frac{2 J_1(k d_A \theta) }{k d_A \theta} \,
,
\end{equation}
where $P_{g\delta}(k)$ is the cross power spectrum between galaxies, $g$, and dark matter, $\delta$. Here, $\da$ is the comoving angular diameter distance and $W^\lens(\rad)$ is the  window function associated with lensing
\begin{eqnarray}
W^\lens(\rad) = \frac{3}{2}\frac{\Omega_m}{a} \left(\frac{H_0}{c}\right)^2 \int_r^{r_0} dr' \frac{\da(\rad)\da(r'-r)}{\da(r')}W^\gal_b(\rad) \, .
\label{eqn:weight}
\end{eqnarray}
Following equation~(\ref{eqn:tangential}) and Guzik \& Seljak (2001), we can write the mean
tangential shear involved with background shear-foreground galaxy distribution as
\begin{equation}
\langle\gamma_t(\theta)\rangle = \int d\rad W^\lens(\rad)W^\gal_f(\rad)
\int \frac{k dk}{2\pi} P_{g\delta}(k) J_2(k d_A \theta) \, .
\label{eqn:tanshear}
\end{equation}

This correlation function has now been measured, and well studied, with Sloan data (Fischer et al. 2000; Guzik \& Seljak 2001, 2002). We will present the first inversion of this correlation function to estimate $P_{g\delta}$ in section~\ref{sec:inversion} of this letter.

\begin{figure}
\centerline{\psfig{file=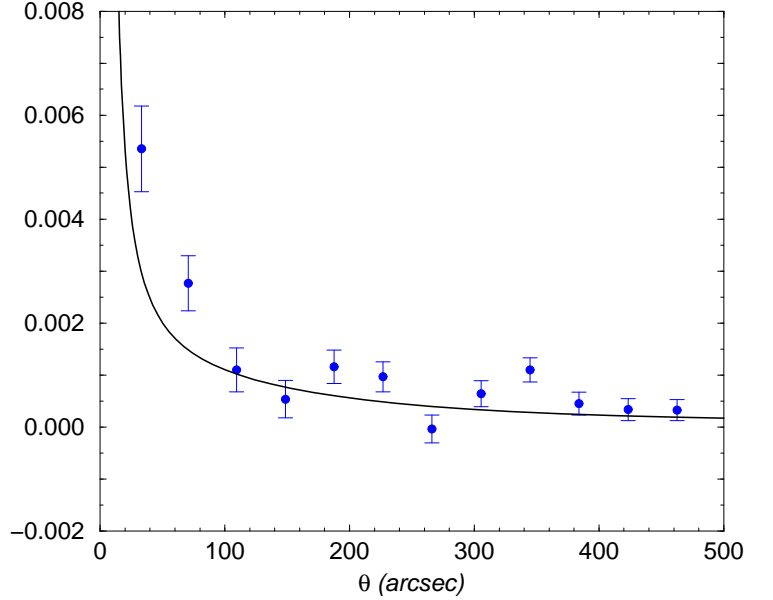,width=3.2in,angle=-90}}
\caption{The foreground galaxy-background shear 
correlation function as measured by the Sloan Digital Sky Survey (Fischer et al. 2000) in the $r'$ band with a magnitude
range of 16 to 18 for foreground galaxies.
 For comparison, we show ahalo model prediction for this correlation with a solid line. In addition to such direct comparisons based on model calculations, the data can be inverted to
estimate the underlying cross power spectrum between galaxies and mass in a model independent manner.} 
\label{fig:sloanshear}
\end{figure}

\begin{figure}
\centerline{\psfig{file=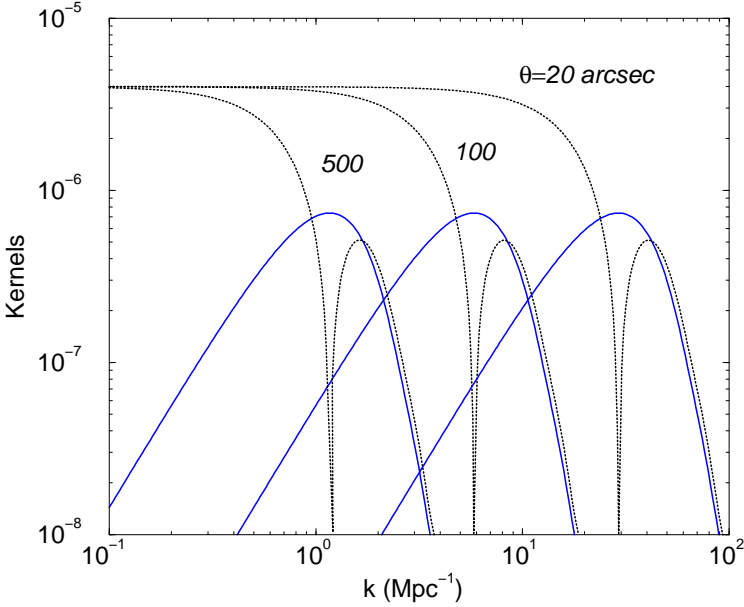,width=3.2in,angle=-90}}
\caption{The kernels involved with the projection of
three dimensional galaxy-mass cross power spectrum to the two dimensional galaxy-shear
correlation (with a $J_2$; solid lines) and the foreground-background galaxy
correlation (with a $J_0$; dotted lines). We show kernels for three fixed values of
angular scales, $\theta$, projected at the same angular diameter distance.
Note that the galaxy-shear correlation
kernel extends to larger wave numbers than  the kernel associated with the galaxy-galaxy correlation. Also, the galaxy-shear correlation
kernel is sharply peaked allowing a better behaved inversion than
in the case with the simple galaxy correlation function.}
\label{fig:j0j2}
\end{figure}

\section{Foreground-background source correlation}
 
The second observational probe of the galaxy-mass correlation function
comes from clustering of background sources around foreground objects.
Here, one constructs a correlation function by simply cross-correlating the surface density of background objects, such as quasars (Dolag \& Bartelmann 1997;
Norman \& Impey 2001) or X-ray sources (Cooray 1999),
with  a sample of foreground galaxies.
The dependence on the correlation comes from the fact that
foreground sources trace the mass density field which can potentially
affect the number counts of background sources
by the weak lensing magnification  effect. 
 
To understand this correlation, we can consider a sample of background
sources whose number counts can be written as
\begin{equation}
N(s) = N_0 s^{-\alpha} \,
\end{equation}
where $s$ is the flux and $\alpha$ is the slope of number counts\footnote{Similarly, we can describe this calculation with counts based on magnitudes instead of flux. In that case, one should replace $\alpha$ with $2.5\alpha_m$ where
$\alpha_m = d log N(m)/dm$; the logarithmic slope of the magnitude counts}.
Due to lensing, when the amplification involved is $\mu$,
one probes to a lower flux limit $s/\mu$ while the total
number of sources are reduced by another factor $\mu$; the latter
results from the decrease in volume such that the total surface brightness
is conserved. Thus, in the presence of lensing,
number counts are modified to
\begin{eqnarray}
N(s) &=& N_0 s^{-\alpha} \mu^{\alpha-1} \, .
\end{eqnarray}

In the limit of weak lensing, as more appropriate for the large scale
structure, $\mu \approx (1+2\kappa)$ where  $\kappa$ is the convergence.
This allows us to write fluctuations in background number counts, $\delta N_b(\bn)$, in the
presence of foreground lensing as (Moessner et al. 1997)
\begin{eqnarray}
\delta N_b(\bn) &=& 2(\alpha_b-1)\kappa(\bn) \nonumber \\
&=& 2(\alpha_b-1)
\int d\rad W^\lens(r)\delta(\bn,\rad) \, ,
\label{eqn:background}
\end{eqnarray}
where the lensing weight function integrates over the
background source population following equation~(\ref{eqn:weight}).
 As before, since foreground galaxies trace the distribution of dark matter responsible for lensing,
we can write the correlation between foreground and
background sources as
\begin{eqnarray}
&&w_{fb}(\theta) =  \nonumber \\ 
&& 2(\alpha_b-1) \int d\rad W^\lens(\rad)W^\gal_f(\rad) 
        \int \frac{k dk}{2 \pi} P_{g\delta}(k) J_0(kd_A\theta) \, , \nonumber \\
\label{eqn:b-f}
\end{eqnarray}
where we have simplified using the Fourier expansion of
equations~(\ref{eqn:background}) and (\ref{eqn:foreground}), and have
introduced, again, the galaxy-mass cross power spectrum.
 
Note that in the case where foreground and background sources are
not distinctively separated in radial coordinates, or equivalently in redshift space,  there may be
an additional correlation resulting from the fact that background
sources trace the same overlapping density field in which foreground
sources are found (Moessner et al. 1997). This leads to a clustering term that is proportional to the
 cross power spectrum between foreground
source sample, galaxies in this case, and the population of background sources. 
This clustering component usually becomes a source of contamination for the
detection of background source-foreground galaxy correlation due to weak
lensing alone.

Returning to lensing aspects of the two correlation functions, it is now clear how they probe the galaxy-mass cross power spectrum:
\begin{eqnarray}
 \left[ \matrix{ \langle \gamma_t \rangle \cr w_{fb}}\right](\theta)
 = \int_0^\infty k\,dk\,P_{g\delta}(k,z=z_m) \left[ \matrix{
K_\gamma(k\theta) \cr
K_\kappa(k \theta)}\right]
\end{eqnarray}
 where kernels, $K_i$, involved with galaxy-shear, $i \equiv \gamma$, and galaxy-source
correlations, $i \equiv \kappa$, are respectively
\begin{eqnarray}
K_\gamma(k\theta) &=& \int \frac{d\rad}{2\pi} W^\lens(\rad)W^\gal(\rad) J_2(k d_A \theta) \nonumber \\
K_\kappa(k\theta) &=& 2(\alpha_b-1) \int \frac{d\rad}{2\pi} W^\lens(\rad)W^\gal(\rad) J_0(k d_A \theta)\, .  \nonumber \\
\label{eq:kernels}
\end{eqnarray}
In addition to a factor of $2(\alpha_b-1)$ in $K_\kappa(k\theta)$, which is an overall normalization, the two kernels are distinctively different due to  differences in Bessel functions involved with the projection; A $J_0$ and a $J_2$ in the galaxy-source and galaxy-shear correlations, respectively. Thus, one expects these two probes to be sensitive to different physical scales, or in Fourier wave-number $k$, of the three dimensional galaxy-mass cross power spectrum at the same projected angular scale, $\theta$.

We illustrate the difference between these two kernels in  figure~\ref{fig:j0j2}. For illustration purposes, we take a redshift distribution of foreground sources of the form $n(z) \propto (z/z_0)^2 \exp[-(z/z_0)^{3/2}]$ with a median redshift, $z_0/1.412$, of 0.2. The background  sources  are taken to be 
at a redshift of 0.5 with $\alpha_b=2$. As shown in figure~\ref{fig:j0j2}, due to the behavior associated with $J_2$, the tangential shear-foreground galaxy correlation function probes smaller scales, or larger $k$'s in terms of wave numbers than the foreground-background correlation which depends on a $J_0$. 

Thus, for reasonable angular scales of interest, the shear-galaxy correlation is more sensitive to non-linear aspects of the galaxy-mass cross power spectrum.
This is also the reason why in the so-called halo models, the one halo 
contribution can be used to describe shear-galaxy correlation adequately.
The two-halo term, which describes the clustering between halos 
and proportional to the linear power spectrum, does not make a significant
contribution to the shear-galaxy correlation even at large angular scales.
The same dependence also makes the shear-galaxy correlation a strong
probe of dark matter halo properties. 
The foreground galaxy-background source correlation, however,
 allows a probe at large and linear scales. This is consistent with the
fact that a simple galaxy-galaxy correlation function involves both large scale
clustering contribution at large angular scales and the non-linear
contribution at small angular scales. 

When combined, shear-galaxy and galaxy-source 
correlations can potentially be used to study the cross power spectrum between galaxies and mass over a wide range in physical scale from linear to the non-linear regime.  We highly recommend a combined study of the two weak lensing probes using the same sample of foreground galaxies and, potentially, the same background source sample from which shear measurements are made and which can be correlated directly with the foreground galaxy distribution.  

Note that the galaxy-mass cross power spectrum can be written as $P_{g\delta}(k) =r(k) b(k) P_{\delta \delta}(k)$, where $P_{\delta \delta}$ is the non-linear dark matter power spectrum. Since the galaxy-galaxy power spectrum is $P_{gg}(k)=b^2(k)P_{\delta \delta}(k)$, one can use a combination of galaxy-shear, galaxy-source and foreground galaxy-foreground galaxy
correlations functions to study galaxy bias, $b(k)$, and its correlation, $r(k)$, with respect to the dark matter density field. This can clearly, and easily, be carried out with current and future wide-field imaging surveys, including the Sloan survey. 

While the correlation function of these galaxies will give information on $P_{gg}$, the lensing probes will provide necessary information on $P_{g\delta}$. Finally, information related to $P_{\delta \delta}(k)$ will come from the background shear-background shear correlation function. The current and
planned surveys from instruments such as the MEGACAM and telescopes such as the Large Synoptic  Survey Telescope (Tyson \& Angel 2001) will allow a detailed study of the 
dark matter power spectrum. We  recommend that
these surveys also consider the cross clustering measurements
 between galaxies and shear in order to extract information on
the galaxy-mass cross power spectrum.

\begin{figure}
\centerline{\psfig{file=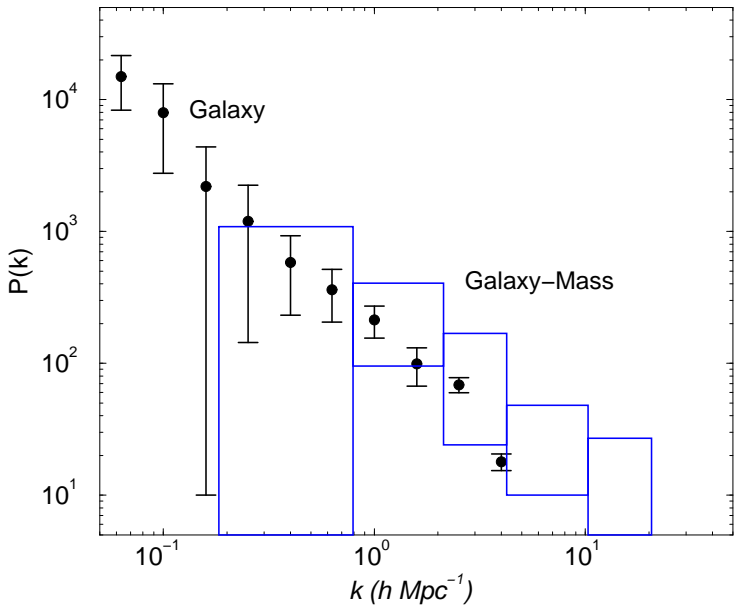,width=3.2in,angle=-90}}
\caption{The galaxy-mass cross power spectrum based on an inversion of the
Sloan's galaxy-shear correlation function (boxes) as measured in Sloan's
r' band.  For comparison, the points with error bars show an estimate for the galaxy power spectrum based on Sloan's two-dimensional clustering data again in the same $r'$ band but restricted to a magnitude range of 19 to 20.}
\label{fig:galaxymass}
\end{figure}

\section{Galaxy-Mass Cross Power}
\label{sec:inversion}

As a first attempt at directly extracting the galaxy-mass cross power spectrum, we can apply above discussion  to
published measurements in the literature. In terms of the
tangential shear-galaxy  correlation, the best results  probably come from the Sloan Digital Sky Survey (Fischer et al. 2000). However, 
no published results are still available for the background-foreground galaxy correlations from Sloan, 
though
initial work towards such a study is currently in progress (Jain et al. in preparation).  Similar work is also in progress to cross correlate galaxy and
quasar samples. Thus, we will only consider
the galaxy-shear correlation as a probe of the galaxy-mass cross power spectrum here.

In figure~\ref{fig:sloanshear}, we compare measurements of this correlation with
a prediction based on the so-called halo model (see, Cooray \& Sheth 2002 for a recent review). 
Here, for simplicity, we take a simple description for galaxy number counts as introduced in Seljak (2000) and
extended later in Guzik \& Seljak (2001).
In calculating the expected
correlation function, we have used the expected redshift distributions for foreground and background galaxies in Sloan samples following
Dodelson et al. (2001) and McKay et al. (2001).  
A more thorough study of the weak lensing
shear-galaxy cross-correlation, under the halo model, is available in
Guzik \& Seljak (2001). We focus on a model independent extraction
of the galaxy-mass cross power spectrum here.

In order to invert the above correlation to estimate the galaxy-mass power spectrum, we follow the approach advocated in Dodelson et al (2001; see, also, Dodelson \& Gazta\~naga 2000). This approach has been applied to 
an inversion  of the Sloan's clustering results, mainly the correlation function, $w(\theta)$, and the angular power spectrum, $C_l$, of galaxies.  We can write the associated inversion equation  as
\begin{equation}
\vec{d} = {\bf K} \vec{P} + \vec{n} \, ,
\label{eq:vector}
\end{equation}
where, in the present case,
 $\vec{d}$ is a vector containing the data related to $\langle \gamma_t(\theta_i)\rangle$
with an associated noise vector $\vec{n} $ and ${\bf K}$ is a matrix containing kernels at each  $\theta_i$ where the correlation function is measured and at each $k_i$ for which cross-power spectrum estimates are desired. The inversion problem involves estimating $\vec{P}$ given
other vectors and the matrix ${\bf K}$. Similar to the notation in Dodelson et al. (2001), the
matrix ${\bf K}$ differs from kernels defined in equation~(\ref{eq:kernels})
due to an additional factor of $k\, dk$.
By appropriately renormalizing equation~(\ref{eq:vector}) with noise,
 following Dodelson et al. (2001), we consider the minimum variance estimate of the galaxy-mass cross power spectrum. 

We refer the reader to Dodelson et al. (2001) for  details on this approach and illustrate our results in figure~\ref{fig:galaxymass}. In the same figure, we also compare our estimate for the galaxy-mass cross power
spectrum with an estimate for the galaxy-galaxy power spectrum 
in Sloan's $r'$ band with magnitudes between 19 and 20. 
While the redshift distribution of galaxies in the shear-galaxy correlation and the galaxy-galaxy correlation may be different, with a carefully selected sample, it is likely that estimates on parameters such as bias and correlations
may eventually be possible. We do not make detailed comparisons between Sloan's galaxy-mass and galaxy-galaxy
power spectra for possible mismatches in redshift as well as mismatches in
 magnitude ranges; While, the
foreground galaxy sample related to the galaxy-mass cross power spectrum is
 between 16 and 18 in the $r'$ band,   estimates 
for the galaxy-galaxy power spectrum have only been considered for
 galaxies with $r' \geq 18$ in Dodelson et al. (2001).

There is, however, one significant difference between the inversion in Dodelson et al. (2001) and the one performed here. The difference, in fact, makes the inversion discussed here more stable and reliable. As suggested, the kernel associated with the shear-galaxy correlation involves a $J_2$ which is better peaked than that associated with a simple galaxy-galaxy
correlation, mainly a $J_0$.  Also, estimates of the power are less correlated while, with a $J_0$, one finds a significant covariance matrix for power spectrum estimates.  The same is also true in the real space; Because of the broadness of the $J_0$, measurements of the galaxy correlation function are correlated (Eisenstein \& Zaldarriaga 2000) while we do not expect this to be the case for the galaxy-shear correlation function. As shown in figure~\ref{fig:j0j2}, another  minor observation is that the kernel associated with 
galaxy-galaxy correlation function, $J_0$, extends to larger $k$ with a decrease in the angular scale of interest.
This makes the simple galaxy-correlation function monotonically increasing with decreasing angular scale. This is not the case with the shear-galaxy correlation. The function only increases at small angular scales due to the additional dependence of $k$ that integrates over kernels shown in figure~\ref{fig:j0j2}.

Even though an inversion of the galaxy-shear correlation is better behaved than the one involving galaxy-galaxy correlation, our estimate for the galaxy-mass power spectrum should be considered as preliminary. We have neglected the covariance, though small, in the angular space between data points. Though the associated  kernel makes such correlations substantially small, a proper accounting of the covariance of measurements is clearly required.  While no knowledge is  available on this aspect, it can eventually be obtained through semi-analytical models such as the one adopted to determine the covariance of galaxy correlation function in Sloan or using data themselves (for e.g., Scranton et al. 2001). 

Also required for this analysis is the precise redshift distribution of background and foreground galaxies. While there is  limited knowledge on this, with photometric data calibrated based on  spectroscopic observations, this is likely to improve in the future. To keep contaminants small, it is important that the two samples are disjoint in the redshift space when estimating 
the foreground galaxy-background source correlation function. We recommend that the measurement of the shear-galaxy correlation be followed in the future with a measurement of the foreground galaxy-background galaxy correlation and a correlation of the foreground galaxies themselves. A combined inversion of these correlation functions will clearly improve our knowledge on the cross clustering between galaxies and mass as well as biasing properties of galaxies themselves. As more details on Sloan's lensing results become public, we will implement such an approach in the future. 

\section{Summary}

We have discussed two probes of the cross power spectrum between galaxies and mass which effectively use two
aspects of weak lensing involving shearing of background images in one case and the magnification of
background images in the other. The two probes, though similar in most aspects, probe
different physical scales of the galaxy-mass cross power spectrum due to a subtle difference in the kernel
which projects three dimensional clustering to the observable two-dimensional angular space.
When combined, effectively with the same sample of foreground galaxies whose redshift distribution is
known a priori, these two probes  allow a study of the galaxy-mass cross power spectrum
 from linear to non-linear scales.  

While the galaxy-mass cross power spectrum provides information on how
galaxies are correlated with mass, the galaxy-galaxy power spectrum provides information on biasing.
Thus, a complete study, which can be easily carried out with imaging data such as from the Sloan survey is to measure the angular correlation of foreground galaxies, and associated lensing correlation functions discussed here.
The three functions can then be inverted in a consistent manner to obtain
much needed knowledge on
galaxy clustering relative to mass. As a first example of such an approach, we have inverted the 
background shear-foreground galaxy correlation function measured by the
 Sloan Digital Sky Survey and have provided a
 first estimate of the cross power spectrum between mass and galaxies at low redshifts as appropriate for Sloan sample of galaxies in the r' band.

\section*{Acknowledgments} 
 
We thank useful discussions with, and help from, 
Scott Dodelson on inversions, Ravi Sheth on halo models and lensing correlation functions,
and Albert Stebbins on Sloan's shear-galaxy correlation.
This work was funded from a senior research fellowship at Caltech 
from the Sherman Fairchild foundation and from a grant from the
US Department of Energy.


\begin{thebibliography}{}

\bibitem[Bartelmann \& Schneider]{BarSch00}
        Bartelmann, M., Schneider, P. 2001, Physics Reports, 340, 291.

\bibitem[Blandford et al]{Blaetal91}
        Blandford, R. D., Saust, A. B., Brainerd, T. G., Villumsen,
J. V. 1991, MNRAS 251, 60

\bibitem[Cooray]{Coo99}
        Cooray, A. R. 1999, A\&A, 348, 673

\bibitem[Cooray \& Sheth]{CooShe02}
        Cooray, A., Sheth, R. K., 2002, to appear in Physics Reports.

\bibitem[Dodelson et al.]{Dodetal00}
Dodelson, S. \& Gazta\~naga, E. 2000, MNRAS, 312, 774

\bibitem[Dodelson et al.]{Dodelat01}
	Dodelson, S., Narayanan, V. K., Tegmark, M. et al. 2001, ApJ in press
(astro-ph/0107421).

\bibitem[Dolag \& Bartelmann]{DogBar97}
	Dolag, K. \& Bartelmann, M. 1997, MNRAS, 292, 446

\bibitem[Eisenstein \& Zaldarriaga]{EisZal01}
	Eisenstein, D. J. \& Zaldarriaga, M. 2001, 546, 2.

\bibitem[Fischer et al]{Fisetal00}
	Fischer, P., McKay, T. A., Sheldon, E. et al. 2000, AJ, 120, 1198.

\bibitem[Guzik \& Seljak]{GuzSel01}
	Guzik, J. \& Seljak, U. 2001, MNRAS, 321, 439.

\bibitem[Guzik \& Seljak]{GuzSel02}
	Guzik, J. \& Seljak, U. 2002, MNRAS submitted (astro-ph/0201448)
	
\bibitem[Kaiser \& Squires]{KaiSqu93}
	Kaiser, N. \& Squires, G.	1993, ApJ, 404, 411

\bibitem[Limber]{Lim54}
        Limber, D. 1954, ApJ, 119, 655

\bibitem[Mckay et al. 2001]{Mac01}
	Mckay, T. A., Sheldon, E. A., Racusin, J. et al. 2001, ApJ submitted (astro-ph/0108013).

\bibitem[Moessner et al]{Moeetal97}
	Moessner, R., Jain, B., \& Villumsen, J. V. 1997, MNRAS, 294, 291
	
\bibitem[Norman \& Impey]{NorImp01}
	Norman, D. J. \& Impey, C. D. 2001, MNRAS, 121, 2392

\bibitem[Scranton]{Scr02}
	Scranton, R. 2002, ApJ submitted (astro-ph/0205517)

\bibitem[Scranton et al.]{Scr01}
	Scranton, R., Johnston, D., Dodelson, S. et al. 2001, ApJ submitted (astro-ph/0107416)

\bibitem[Seljak]{Sel01}
	Seljak, U. 2000, MNRAS, 318, 203.	

\bibitem[Tyson \& Angel]{TysAng01}
	Tyson, A. \& Angel, R. 2001, In
The New Era of Wide Field Astronomy, ASP Conference Series, Vol. 232. Eds. Roger Clowes, Andrew Adamson, and Gordon Bromage (ASP: San Francisco).




 \end{thebibliography}
\end{document}